\documentclass[showpacs,preprintnumbers,aps,prb,preprint]{revtex4}
\usepackage[dvips,xdvi]{graphicx}
\usepackage{epsfig}

\begin {document}

\title{Resonant two-photon excitation of 1s paraexcitons in Cuprous Oxide}
\author{Yingmei Liu, David Snoke}
\affiliation{Department of Physics and Astronomy\\ University of
Pittsburgh, Pittsburgh, PA 15260}
\date{}
\begin{abstract}
We have created paraexcitons in Cu$_{2}$O via resonant two-photon
generation, and examined their population dynamics by means of
time-correlated single photon detection. Confining the excitons to
a constant volume in a harmonic-potential trap made with
inhomogeneous applied stress along the [001] axis, we find that
paraexcitons are created directly, and orthoexcitons appear
primarily through a up-conversion of the paraexcitons. Hot
excitons are also created via a three-photon process when the IR
laser is non-resonant. Also we generate excitons with two
colliding pulses, and the luminescence is weaker than that from
one beam excitation with same total laser power. These results
show that resonant one-beam two-photon generation of paraexcitons
is a promising way to pursue Bose-Einstein condensation of
paraexcitons.
\end{abstract}
\maketitle

\newpage
\section{Introduction} Naturally-grown high-quality Cu$_2$O is a good candidate for
studying excitonic Bose-Einstein condensation. First, an inversion
symmetry in Cu$_2$O forbids any direct dipole recombination of
excitons, which gives the exciton a long lifetime up to
microseconds. Second, Cu$_2$O is a direct-gap semiconductor with
repulsive exciton-exciton interactions, which means the
electron-hole liquid (EHL) is not stable \cite{bec}. Moreover, the
large binding energy of the excitons (150 meV) is equivalent to a
temperature of 1740 Kelvin, which allows excitons to exist at room
temperature \cite{auger}. The lowest excitonic state is the
1s-``yellow" exciton series, consisting of electrons in the
$\Gamma_6^+$ level of the conduction band and holes in the
$\Gamma_7^+$ level of the valence band with 2.173 eV gap energy.
By electron-hole exchange, the ``yellow" excitons split into a
triplet orthoexciton and a singlet paraexciton, which lies 12 meV
lower \cite{auger}.

Most studies \cite{auger, negoita, snoke} have concentrated on the
higher-lying orthoexciton state since it can be easily created by
a single-photon phonon-assisted absorption. In the single-photon
phonon-assisted excitation process, however, a hot optical phonon
is created which can heat the exciton gas, thereby deterring BEC.
In the past few years, several groups \cite{naka1, kono, kubouchi,
jolk, sun, shen, naka2,naka3} have tried resonant two-photon
excitation of the orthoexciton state.  A major advantage is that
the scattered two-photon laser light will not damage imaging
systems because it is far away from the exciton luminescence
lines, which allows a measurement during the laser pulse. It is
also expected that two colliding pulses in opposite directions
will create excitons directly in the $\textbf{$\vec{k}$}=0$ state,
where a condensate should appear, instead of in a state with
finite momentum.

Resonant excitation of orthoexcitons still has the problem that
the orthoexcitons will convert down into paraexcitons by a phonon
emission \cite{bec}, and the emitted phonon will cause a local
heating, which may prevent BEC. Our present work aims to solve the
above problem with a resonant two-photon excitation of the yellow
paraexciton state, because it is the lowest exciton state and
primarily only decays by direct single-photon emission.  We create
excitons in a harmonic potential trap made with an external
stressor \cite{auger, snoke, trauernicht}, which confines excitons
in a defined volume and changes the local symmetry of Cu$_2$O from
O$_{h}$ to D$_{4h}$. One problem which becomes more important in
the potential trap is the Auger recombination process, which
happens when two excitons collide and one exciton ionizes, taking
the energy of the other exciton that recombined. Therefore the
Auger process not only gives a severe limit in exciton density but
also heats up the exciton gas, which will tend to prevent BEC. The
Auger recombination rate increases roughly as the square of the
stress, with much smaller Auger recombination at the stress lower
than 1.5 kbar \cite{auger, temperature}. However, because the
phonon-assisted orthoexciton and direct paraexciton luminescence
lines are separated at high stress but substantially overlapped at
low stress \cite{trauernicht}, we have to use high stress to
separate the luminescence lines. In this work, the inhomogeneous
applied stress is kept at 1.9 kbar, a trade-off point which is
good enough to spectrally separate the paraexciton line and
provides a relatively small Auger process.

In this paper, we present experimental results for two
experimental configurations, with one IR laser beam and with two
colliding IR laser pulses. We demonstrate the resonant creation of
paraexcitons under two-photon excitations with several strong
evidences. Then we propose a model of the excitation mechanism for
the paraexciton and orthoexciton in the two-photon excitations,
based on our study on selection rules and polarization dependence
of the yellow excitons from group theory.

\section{Experimental set-up}
The Cu$_2$O sample, a $4.9\times3\times3$ mm bulk crystal, is
immersed in the liquid helium in a Janis optical cryostat, at a
temperature controlled by a homemade PID temperature controller.
The lowest temperature in the cryostat is 1.6 Kelvin. As shown in
Fig. 1, in one-beam experiments the sample is oriented as follows:
one [110] surface is illuminated by an IR laser beam and the
exciton luminescence is collected from the perpendicular
[$\bar{1}$10] surface by an imaging lens which focuses the light
onto the entrance slit of a $\frac{1}{4}$-meter spectrometer, with
a choice of a CCD camera and a photo-multiplier tube (PMT) as
detectors. The exciton population dynamics are examined by means
of time-correlated single photon detection. We can also create
excitons by splitting the IR laser beam into two
counter-propagating parts which meet in the Cu$_{2}$O sample at
the same time from opposite [110] directions, with a controlled
time delay. The repetition rate, the duration and the spectral
width of the IR laser are 250 KHz, 200 fs and 20 nm, respectively.
In order to compare, we also do experiments under single-photon
excitation with a cavity-dumped 3.8 MHz and 50 ps red dye-laser
tuned to the exciton resonance for the same crystal orientation.

Excitons are confined by a harmonic potential trap made with a 6
mm radius curved glass stressor applied along the [001] axis from
the top surface of the Cu$_{2}$O crystal. Under the [001] stress,
we find strong paraexciton signal, although Naka and Nagasawa
\cite{naka1} reported that paraexcitons were too weak to be
detected. The main reason is that our femto-second laser system
has much greater instantaneous intensity than their nano-second
laser. Moreover, the exciton luminescence is collected from a
direction perpendicular to the incident laser beam in this work,
which has several advantages over their forward scattering
geometry \cite{naka1,naka2,naka3}. For example, the second
harmonic scattering of the incident infrared laser light does not
mix with the exciton luminescence.

\section{One-beam, two-photon excitation}
\subsection{Two-photon or three-photon process?}
When we create excitons with an IR laser, the first question which
arises is how we can say it is not a three-photon or four-photon
excitation, but a two-photon excitation. One way to answer this
question is from the dependence of the exciton luminescence
intensity on the IR laser power, that is to say, the excitation is
a n-photon process if the intensity of exciton luminescence is
proportional to (laser-power)$^{n}$. We performed a series of
experiments for a broad wavelength range of IR lasers, with an
inhomogeneous stress of 1.9 kbar along the [001] direction and the
paraexciton line position at 614.1 nm. As shown in Fig. 2, our
experimental results indicate that the excitation is a two-photon
process when the IR laser is at the paraexciton resonant position,
and it is a three-photon process under off-resonant IR laser, for
example at 1240 nm.  All experiments in the following discussions
are under resonant two-photon excitation of the paraexciton state.
We note that even at this wavelength, the three-photon process
will be responsible for about 25\% of the total exciton creation
at the highest laser powers we use, and it will dominate at high
excitation power, which may also create another barrier to BEC.

\subsection{Strong evidence for resonant paraexciton excitations}
With an IR laser tuned to one half the paraexciton ground-state
energy, several experimental results convince us that we directly
create paraexcitons, and orthoexcitons primarily come from the
excitonic Auger process. The first strong evidence is shown in
Fig. 3, the time-integrated luminescence intensity as a function
of IR laser power from 55 mW to 3 mW. As mentioned above, under
the 1.9 kbar stress along the [001] direction, the direct
paraexciton line is at 614.1 nm. Although paraexcitons primarily
only decay by single-photon emission, there are two emission lines
for orthoexcitons, as shown in Fig. 3, the direct orthoexciton
luminescence line at 611.7 nm, which corresponds to emission of a
single photon, and phonon-assisted orthoexciton luminescence line
at 615.7 nm, which corresponds to emission of a single photon and
a $\Gamma^{-}_{12}$ optical phonon with energy of 13.8 meV. There
is only paraexciton luminescence when the laser power is very low,
indicating that the resonant IR laser creates the paraexciton
directly. Orthoexcitons start to appear at higher laser power,
around 5 mW as seen in Fig. 3, and become stronger and stronger
relative to the paraexcitons with increasing the laser power,
which implies that orthoexcitons are created not directly by the
laser, but by a density-dependent process from paraexcitons.

A second evidence is found in time-resolved paraexciton
luminescence data. Fig. 4 shows paraexciton luminescence intensity
as a function of time after laser pulses in two different
pulse-laser excitations. In one case, two-photon excitation
resonant with the paraexciton state is used, and in the second
case, single-photon excitation tuned to the bottom of the
orthoexciton phonon-assisted absorption is used. The average
powers of the IR laser and the red laser used in the excitations
were adjusted to get the same total integrated exciton
luminescence intensity. These average powers were 70 mW and 0.9
mW, respectively. There are several similarities between the two
cases. The paraexciton luminescence intensity in both cases
reaches a steady-state value long after the laser pulses which
indicates lifetimes of paraexcitons longer than the 260 ns period
between the laser pulses, and the paraexciton luminescence
intensities at time just before the laser pulses are comparable,
equal to the steady-state value left from the previous laser
pulse. However, in the case of resonant two-photon excitation, the
paraexcitons reach maximum intensity immediately after the laser
pulse, consistent with the paraexcitons being created directly by
the laser. In the case of single-photon excitation of
orthoexcitons, paraexcitons do not appear right after the laser
pulse, but are slowly created in a time interval of more than 10
ns from a process which is known to be phonon-assisted
orthoexciton down-conversion \cite{auger}. The rise time of 10 ns
is longer than the reported orthoexciton-paraexciton conversion
time of 4 ns, \cite{auger} presumably because the paraexcitons
must also cool down before they appear in the single-photon
luminescence, which involves only states near the band bottom.

Another strong evidence is the power dependence of time-resolved
data for both paraexcitons and orthoexcitons under the two-photon
excitation resonant with the paraexciton state. Fig. 5 shows two
important things. First, paraexcitons appear right after the laser
pulse, but orthoexcitons are created slowly after a few
nanoseconds. Second, orthoexcitons are created more slowly at lower IR
laser power, which again indicates a density-dependent process for the
creation of orthoexcitons.

All the above results lead us to a model in which paraexcitons are
created directly by IR laser pulses resonant with the paraexciton
state, then paraexcitons convert to orthoexcitons through a
density-dependent process which may be the well-known excitonic
Auger process \cite{auger,snoke, temperature}. In each Auger
process, two excitons collide and end up with one exciton
recombining and the other exciton ionizing. Because the
orthoexciton is a triplet state and the paraexciton is a singlet
state, and spin is randomly selected in ionization, 75 percent of
the ionized excitons in the Auger process will be returned as
orthoexcitons and 25 percent of these excitons will be returned as
paraexcitons.  This implies that at high excitation power, the
density of orthoexcitons will exceed that of paraexcitons after a
few nanoseconds, consistent with the results in Fig. 5. The
observation that the creation of orthoexcitons becomes slower at
lower exciton density is consistent with the fact that the
density-dependent Auger process is slower at lower exciton
density.

\subsection{Selection rules and polarization dependence in two-photon
excitations} The laser pulses we used for the two-photon
excitation have spectral width of around 20 nm. One would
therefore expect that if the orthoexciton and paraexciton cross
sections for two-photon absorption are comparable, that we should
then see a significant direct creation of orthoexcitons in
addition to paraexcitons, even when the laser is tuned to the
paraexciton resonance. The results discussed above indicate that
primarily only paraexcitons are being created, however. Is this
consistent with the selection rules? Here we take a close look at
selection rules of paraexcitons and orthoexcitons in two-photon
excitation.

For our one-beam two-photon excitation, the two input beams have
the same energy, $\omega_1=\omega_2=\omega$, and the same
polarization, $\lambda_1=\lambda_2=\lambda$. Therefore, the
absorption coefficient in the dipole transition is proportional
to:
\begin{equation}
\sum_j|\sum_i{\frac{\langle\Gamma_{ex}\mid\textbf{$D_j$}\mid{i}\rangle\langle
i\mid\textbf{$D_j$}\mid{\Gamma_{1}^+}\rangle}{E_i-\hbar\omega}|^2},
\end{equation}
where $\mid{i}\rangle$ are the intermediate states,
$\Gamma_{1}^{+}$ and $\Gamma_{ex}$ are the vacuum state and the
1s-``yellow" exciton state respectively.
$\textbf{$D_j$}=\hat{\lambda}\cdot{p_j}$ are possible dipole
operators for a certain input polarization $\lambda$, where $p_j$
are operators $\hat{x}$, $\hat{y}$ and $\hat{z}$ in three
dimensions.

Because the dipole operator is $\Gamma_{5}^{-}$ in the $D_{4h}$
group \cite{symmetry} for the stressed Cu$_2$O, all allowed
internal states should have the negative parity, which could be
the $p$-states in the ``yellow" exciton series, the ``blue" and
``indigo" exciton series \cite{symmetry}. By taking into account
the energy difference between the ``yellow", ``blue" and ``indigo"
excitons, the dominant internal states are the 2$p$-``yellow"
excitons with $\Gamma_{1}^{-}$, $\Gamma_{2}^{-}$,
$\Gamma_{3}^{-}$, $\Gamma_{4}^{-}$ and $\Gamma_{5}^{-}$
symmetries.
\begin{table}[!h]
\begin{tabular}{|c|c|c|}
\hline
 $D_{4h}$ Group&Symmetry&Basis\\
\hline 2$p$-``yellow" excitons&${}^{1}\Gamma_1^-$&$(x^2-y^2)xyz$\\
&${}^{1}\Gamma_2^-$&$z$\\
&$2({}^{1}\Gamma_3^-)$&$xyz$\\
&$2({}^{1}\Gamma_4^-)$&$(x^2-y^2)z$\\
&$3({}^{2}\Gamma_5^-)$&$x, y$\\
\hline 2$s$-``yellow" excitons&${}^{2}\Gamma_5^+$&$yz, zx$\\
&${}^{1}\Gamma_4^+$&$xy$\\
&${}^{1}\Gamma_3^+$&$(x^2-y^2)$\\
\hline
\end{tabular}
\caption{\small{The symmetry properties of the intermediate states
for both the electric dipole and quadrupole operators in TPE under
the stress along the [001] axis \cite{symmetry,group}.}}
\end{table}

To get a general polarization dependence for excitons in the
dipole transition, we substitute the three possible dipole
operators and possible symmetries of the 2$p$-``yellow" exciton
states into Equation (1). The calculation result, the polarization
dependence for the para and ortho in the dipole transition matrix
element under stress along the [001] axis, is explicitly listed in
Table 2.

In the quadrupole transition, the absorption coefficient can be
calculated in the same way, which is proportional to:
\begin{equation}
\sum_j|\sum_i{\frac{\langle\Gamma_{ex}\mid\textbf{$Q_j$}\mid{i}\rangle\langle
i\mid\textbf{$Q_j$}\mid{\Gamma_{1}^+}\rangle}{E_i-\hbar\omega}|^2},
\end{equation}
where $\mid{i}\rangle$ are the intermediate states,
$\Gamma_{1}^{+}$ and $\Gamma_{ex}$ are the vacuum state and the
1s-``yellow" exciton state respectively. $\textbf{$Q_j$}$ are
possible quadrupole operators for a certain input polarization
$\lambda$, which are expressed as the following \cite{group}:
\begin{eqnarray}
\lefteqn{ \quad Q = \sum_{i,j}k_i\lambda_jr_ir_j}
\nonumber\\
&&\Rightarrow Q =
k_x\lambda_xx^2+k_x\lambda_y(xy)+k_x\lambda_z(xz)+k_y\lambda_x(yx)+\nonumber\\&&\quad\quad\quad
+k_y\lambda_yy^2+k_y\lambda_z(yz)+k_z\lambda_x(zx)+\nonumber\\&&\quad\quad\quad+k_z\lambda_y(zy)+
k_z\lambda_zz^2 \end{eqnarray} where \textbf{k} and
\textbf{$\lambda$} are wave vector and polarization vector of the
input IR beam.

It is not possible to have a mixed dipole-quadrupole transition
because the two operators have different parities, and there is no
allowed internal state for the mixed transition.

There are five possible quadrupole operators in the $D_{4h}$
group, ${}^2\Gamma_{5}^{+}$, ${}^1\Gamma_{1}^{+}$,
${}^1\Gamma_{3}^{+}$ and ${}^1\Gamma_{4}^{+}$, as listed in Table
1. Because the five quadrupole operators have positive parity, all
neighboring exciton states with positive parity can be the
intermediate states for the two-photon quadrupole transition.
However, the dominant internal states are the 2$s$-``yellow"
exciton states, which are the closest neighbor for paraexcitons
and orthoexcitons in energy. The polarization dependence of
orthoexcitons and paraexcitons are calculated from Equation (2)
with the five possible quadrupole operators and four
2$s$-``yellow" exciton states, as listed in Table 2.

\begin{table}[ht!]
\begin{tabular}{|c|c|c|c|c|}
\hline Symmetry group&Transition&Wave vector \textbf{k}
&Orthoexciton&Paraexciton\\
\hline
O$_{h}$&Dipole&(1, 0, 0)& $\sin^22\theta$&Forbidden\\
&&(1, 1, 0)&$\sin^22\theta+\cos^4\theta$&Forbidden\\
&&(1, 1, 1)& 1&Forbidden\\
\hline
O$_{h}$&Quadrupole&(1, 0, 0)& $\sin^22\theta$&Forbidden\\
&&(1, 1, 0)&$(1+a\cos2\theta)\sin^2\theta$&Forbidden\\
&&(1, 1,
1)&\quad$(1+a\cos2\theta+b\cos4\theta)$&Forbidden\\
\hline
D$_{4h}$&Dipole&(1, 0, 0)& $\sin^22\theta$&$\cos^4\theta$\\
&&(1, 1, 0)&$\sin^22\theta+\cos^4\theta$&Forbidden\\
&&(1, 1, 1)& 1&$\sin^22\theta$\\
\hline
D$_{4h}$&Quadrupole&(1, 0, 0)& $\sin^22\theta$&$\sin^4\theta$\\
&&(1, 1, 0)&$(1+a\cos2\theta)\sin^2\theta$&Forbidden\\
&&(1, 1, 1)&\quad$(1+a\cos2\theta+b\cos4\theta)$&$\sin^22\theta$\\
\hline
D$_{2h}$&Dipole&(1, 0, 0)&$\sin^22\theta$&$(a\cos^2\theta+b\sin^2\theta)^2$\\
&&(1, 1, 0)&$\sin^22\theta+\cos^4\theta$&$(a\cos^2\theta+b\sin^2\theta)^2$\\
&&(1, 1, 1)&1&\quad$(1+a\cos2\theta+b\sin2\theta)^2$\\
\hline
\end{tabular}
\caption{\small{Polarization dependence of orthoexcitons and
paraexcitons in the dipole and quadrupole transitions for various
symmetries in two-photon excitation, where $a$ and $b$ are
constant factors which depend on the matrix elements to the
internal states. $\theta$ is the angle between the horizontal
polarization and the polarization of the input IR beam.}}
\end{table}

According to Table 2, when there is no external stress, the
two-photon transition is only allowed for orthoexcitons with
polarization dependence of $\sin^22\theta+\cos^4{\theta}$ in the
dipole transition, which is confirmed by experimental reports from
Kono \emph{et al.} \cite{kono} and Kubouchi \emph{et al.}
\cite{kubouchi}

From the same table, for our [110] laser incident direction, if
the stressed crystal is in the D$_{4h}$ group, paraexcitons should
be forbidden in both dipole and quadrupole two-photon transitions,
which is consistent with the prediction of Inoue-Toyozawa
\cite{toyozawa}, Bader-Gold \cite{gold} and Ablova-Bobrysheva
\cite{ablova}. $\theta$ is the angle between the polarization of
the input laser beam and the horizontal polarization. However, our
experimental results in Figure 6 show that there is strong
paraexciton luminescence, and the luminescence from both
paraexcitons and orthoexcitons has the same intensity for the
vertical and horizontal polarizations. This can be explained due
to the reduction of the symmetry. That is to say, the crystal is
in the D$_{2h}$ group for most of the harmonic potential well,
although the crystal is in the D$_{4h}$ symmetry group along the
axis of the stress. In the D$_{2h}$ group, the polarization
dependence of excitons in the direct dipole transition calculated
from Equation 1 is listed in Table 2. For our [110] laser incident
direction in the D$_{2h}$ group, the paraexciton follows a
polarization dependence of $(a\cos^2\theta+b\sin^2\theta)^2$,
where $a$ and $b$ are constant factors. Therefore, if $a=b$, there
is no detectable difference between the vertical and horizontal
polarization. Because the circular polarization is just a
combination of both the vertical and horizontal polarization, we
don't expect to see any difference in the exciton
photoluminescence intensity when the input laser is changed to the
circular polarization from either vertical or horizontal
polarization, which is well confirmed by the experimental results
in Figure 6.

The orthoexcitons following the exact polarization dependence of
the paraexcitons is another good evidence for our proposed model
in the resonant two-photon excitation: paraexcitons are created
directly by the IR laser and orthoexcitons are created indirectly
through a up-conversion of the paraexcitons.

Moreover, we also note that the birefringence changes the
polarization of the incident infrared laser beam and the changes
are different at different positions in the harmonic potential
well. In general, there is substantial polarization mixing for the
whole harmonic trap.

At this point, one may ask why the paraexciton dipole and
quadrupole matrix elements should dominate the creation of
excitons in our two-photon excitation, since the orthoexciton
transitions are allowed. One reason is that the IR laser is tuned
to the resonant wavelength of the paraexciton state in this work,
which gives an on-resonant two-photon excitation for paraexcitons
and an off-resonant three-photon excitation for orthoexcitons.
Moreover, all our experiments except Figure 6 are done with a
vertically polarized IR laser, that is to say,
$\theta=90^{\small{0}}$ in Table 2, which means that the
orthoexciton polarization dependence for our [110] laser incident
direction in the D$_{2h}$ group is
$\cos^4{\theta}+\sin^2{2\theta}=\cos^4{(90^{\small{0}})}+\sin^2{(180^{\small{0}})}=0$.
Therefore, the orthoexciton dipole matrix element is forbidden for
most of this work. At $\theta=0^{\small{0}}$, the horizontal
polarization, we would expect some orthoexcitons created directly
by the IR laser, but the matrix element may be smaller.

Therefore, we can explicitly express our results as the following:
confined by a harmonic potential trap created with an external
stress, when an IR laser pulse tuned to one half the paraexciton
ground-state energy is shining along the [110] crystalline
direction of the Cu$_{2}$O sample, paraexcitons are directly
created by the IR laser pulse, and orthoexcitons are created
primarily from the paraexciton up-conversation process at the
early time after the laser pulse.

\subsection{Is there BEC of paraexcitons?}
Since we can resonantly create the paraexciton via resonant
two-photon excitation, should we see BEC of the paraexcitons? Our
estimate of the density of paraexcitons indicates that we should
not expect to see this. Fig. 7 shows an integrated spatial profile
of the paraexciton in the Cu$_{2}$O crystal at 1.6 Kelvin, which
gives the paraexciton excitation volume V = $\frac{4}{3}\pi(20 \mu
m)^{3}$. As mentioned in Section (3.B), when detected by PMT, the
250 KHz resonant IR laser with average power of 70 mW creates the
same exciton luminescence intensity via two-photon excitation as
the 3.8 MHz dye-laser at 607.5 nm with 0.9 mW average power does
via single-photon excitation. If we assume that 5\% of the 0.9 mW
average power is absorbed in the single-photon excitation
\cite{negoita2}, that one absorbed photon creates one exciton in
the single-photon excitation, and that comparable luminescence
intensity implies comparable exciton density, then we find the
density of excitons in the two-photon excitation is
\begin{equation}
n = \frac{3.8 MHz}{250 KHz}*\frac{0.9 mW*5\%}{3.8
MHz*\frac{2\pi\hbar c}{\lambda}*V} = 1.6*10^{16} cm^{-3}.
\end{equation}
where $c$ is the speed of light, $\frac{2\pi\hbar c}{\lambda}$ is
a single photon energy in the single-photon excitation with a
laser at 607.5 nm, and the ratio $\frac{3.8 MHz}{250 KHz}$ is the
calibration of the photon-counting efficiencies in the
time-correlated single photon detection for the two laser systems
with different repetition rates. This corresponds to $0.06\%$
absorption of the infrared laser light.

To calculate the critical exciton density for BEC, we have to know
the exciton temperature, which can be calculated in two ways. One
way is to fit high temperature tail of the paraexciton
luminescence with a Bose distribution function,
$\varepsilon^{1/2}/(e^{\varepsilon/(k_{B}T)}-1)$, where
$\varepsilon$ is paraexciton energy. The other way is to find the
full width at half maximum of the phonon-assisted luminescence
line, which is equal to 3.4$k_{B}T$ in a three-dimensional
harmonic potential trap \cite{auger}. Our calculations from the
above two ways indicate that the exciton has a minimum temperature
of 7 Kelvin, which means the critical exciton density should be
$5*10^{17} cm^{-3}$. \cite{bec} Since the created paraexciton
density is about thirty times less than the required density for
BEC, we do not expect to establish BEC of paraexcitons even with
the maximum laser power from our laser system. With higher laser
power or lower temperature, however, the paraexciton critical
density can be approached with resonant two-photon excitation
which creates paraexcitons directly.

In order to know whether the density-dependent Auger process is
the only paraexciton up-conversion process, we can do an
interesting comparison between Figure 5 and our former work
\cite{auger, temperature}. The former work gives the Auger
constant of $3*10^{-17}\mbox{cm}^3/\mbox{ns}$ around 1.9 kbar,
which implies a decay rate of 2 ns and 4 ns for the input IR laser
power of 90 mW and 45 mW, respectively. While Figure 5 gives
different decay rates, 3 ns for the 90 mW laser and 10 ns for the
45 mW. There are several reasons for the big inconsistency. First,
there may be other up-conversion processes for the paraexciton
besides the density-dependent Auger process. Moreover, the
paraexcitons must also cool down before they appear in the
single-photon luminescence.

\section{Creating excitons with two colliding pulses}
Besides doing experiments with one-beam, two-photon excitation
resonant with the paraexciton state, we can excite excitons with
two colliding pulses, that is to say, split the IR laser beam into
two parts and shine them into the Cu$_{2}$O sample, such that both
pulses reach the exciton stress trap at the same time. In this
case, the $\vec{k}=0$ state is directly excited if the two parts
travel in opposite directions. Therefore, the created excitons
have zero momentum, \textbf{$\vec{k}$} = 0, which means the ground
state is directly excited, where the condensate should appear, if
one photon is from each pulse.

The experimental set-up is shown in Fig. 1. In order to make the
two beams meet at the same time in the Cu$_{2}$O sample, we added
a two-mirror delay system controlled by a precise translation
stage. Fig. 8 shows the dependence of the exciton luminescence
intensity on the time delay between the two colliding pulses. A
comparison between one-beam and two-beam resonant excitations with
the same total laser power, 90 mW, is shown in Fig. 9. The exciton
luminescence in the one-beam excitation is brighter than that of
the two-beam excitation, and creation of orthoexcitons in the
one-beam excitation is faster than with the two-beam excitation,
which indicates more excitons are created in the one-beam
excitation.

To explain this phenomenon, one must consider the dependence of
the density of states on the momentum \textbf{$\vec{k}$}. In the
case of two photons travelling in the same direction, the excitons
are created in the region of the (weak) polariton mixing.  Two
photons travelling in opposite directions will create an exciton
at ${\vec{k}} =0$, where the density of states is much lower. In
the case of two-beam excitation, there will also be one-beam
excitation with half of the total power of the two-beam
excitation. Since the exciton density in the two-photon process is
proportional to the square of laser power, the one-beam excitation
with half of the total power will have one fourth the efficiency.
The sum of all processes in the two-beam excitation ends up being
less than the total in the one-beam case.

\section{Conclusion}
We resonantly create paraexcitons with an IR laser beam tuned to
one half the paraexciton ground-state energy. This is surprising,
since it is generally assumed that the cross section for
paraexciton creation is always much less than that for
orthoexciton. Our experimental results are all consistent with
this result, however, including the polarization dependence and
time-dependent data. The exciton creation efficiency in resonant
two-photon excitation is greater for one-beam excitation than for
two colliding pulses, but the colliding pulse method may be useful
for direct creation of a condensate in the ground state. At
present, the paraexciton density in this work is thirty times less
than the required density for BEC of paraexcitons, but with higher
laser power from stronger IR laser sources or lower temperature,
the critical density can be approached with one-beam two-photon
excitation resonant with the paraexciton state.

{\bf Acknowledgements}. This work has been supported by CRDF-MRDA
Award No. MP2-3026 and NSF Award No. DMR-0102457. Samples of
Cu$_2$O were obtained from P. J. Dunn of the Smithsonian
Institute. We would like to thank Robert P. Devaty for discussions
on group theory, Albert Heberle for suggestions on spectral
filtering, M. Kuwata-Gonokami of Tokoyo university on the
selection rules and thank Vincenzo Savona for helpful
conversations on the two-beam, two-photon excitation.

\newpage
\begin{figure}[!ht]
\begin{center}
\includegraphics[width=7in,height=5in]{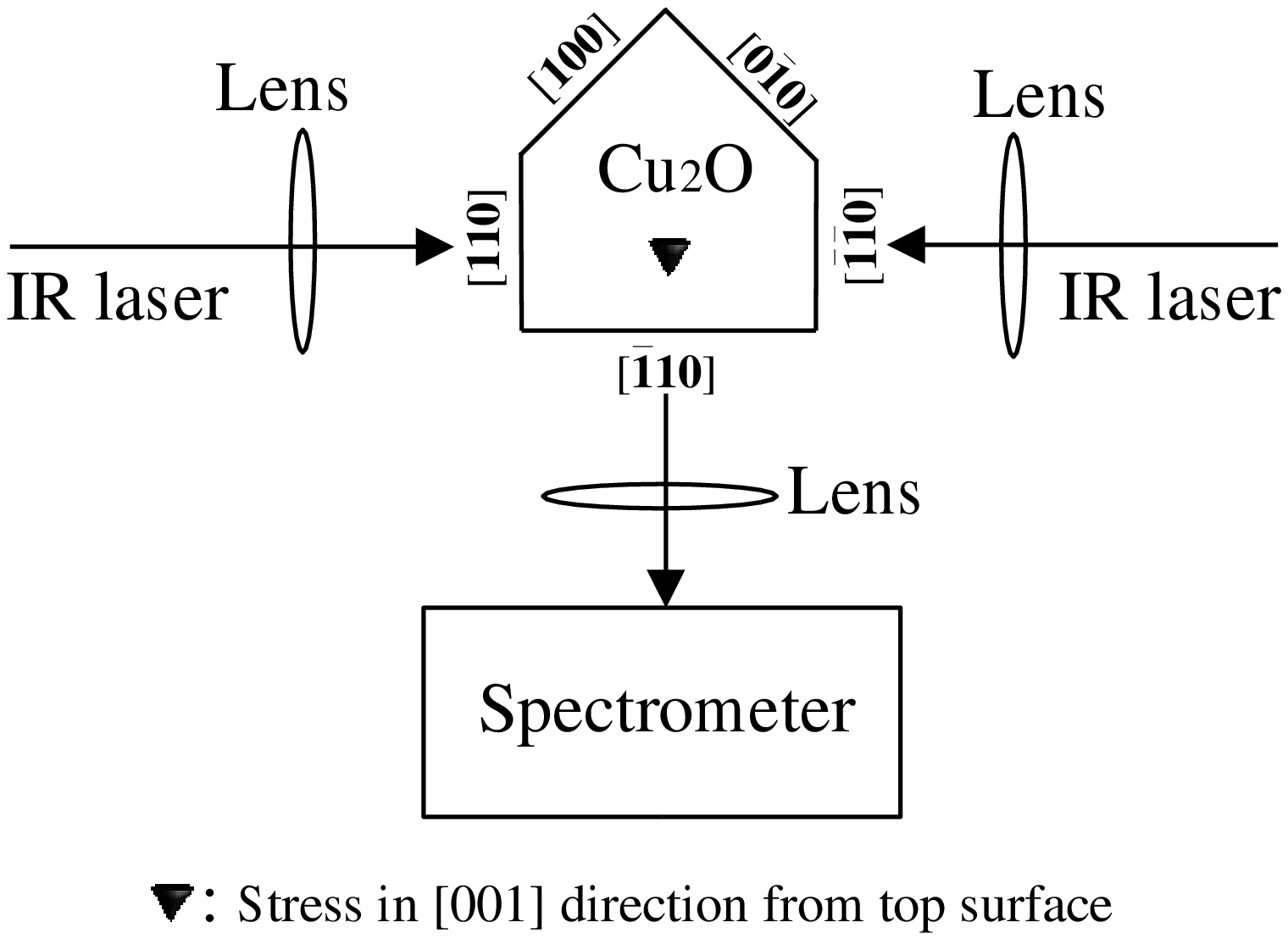}
\caption{\label{fig1.fig} Schematic of the experimental set-up in
this work.}
\end{center}
\end{figure}

\begin{figure}[!ht]
\begin{center}
\includegraphics[width=7in,height=5in]{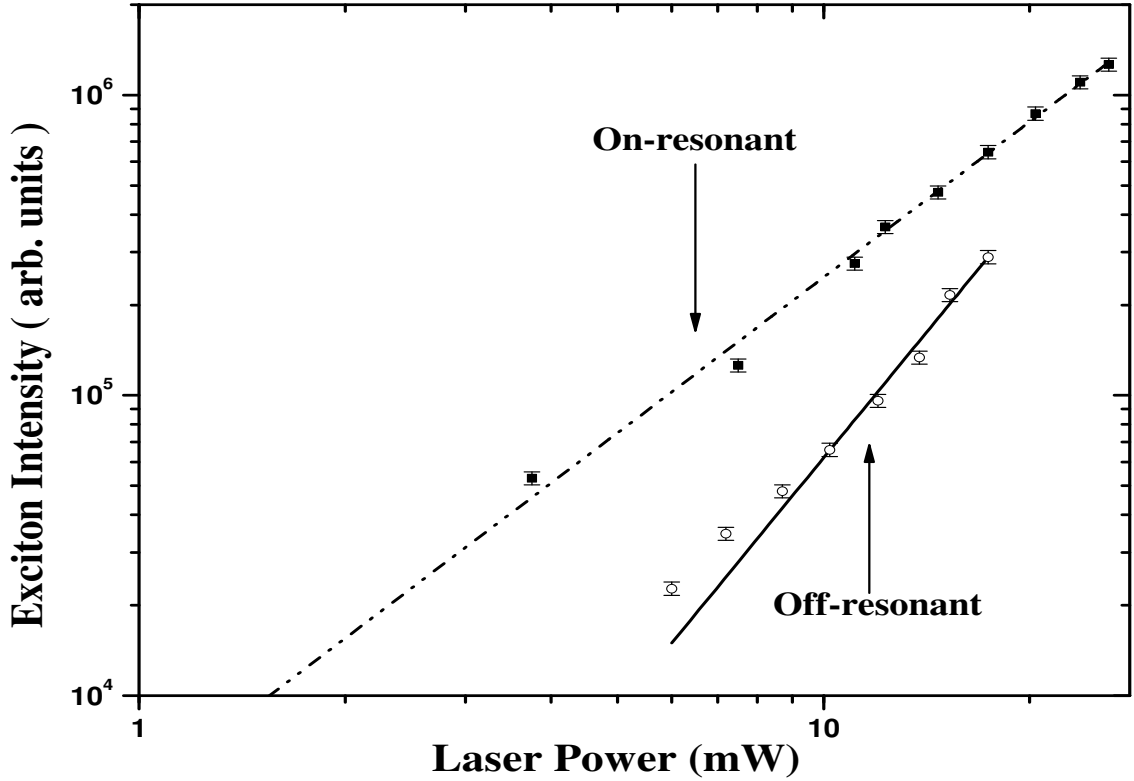}
\caption{\label{fig2.fig} When the IR laser is at 1228 nm,
resonant with the paraexciton state, integrated exciton
luminescence intensity (black squares in the figure) can be best
fit to $P^{1.7\pm0.1}$ (black dashed line), where $P$ is the input
laser power. This is consistent with a two-photon process within
the experimental error. If the laser is off-resonant, at 1240 nm,
the integrated exciton luminescence intensity (black open circles)
can be best fit to $P^{2.8\pm0.2}$ (black solid line in the
figure), consistent with a three-photon process. }
\end{center}
\end{figure}

\begin{figure}[!ht]
\begin{center}
\includegraphics[width=7in,height=5in]{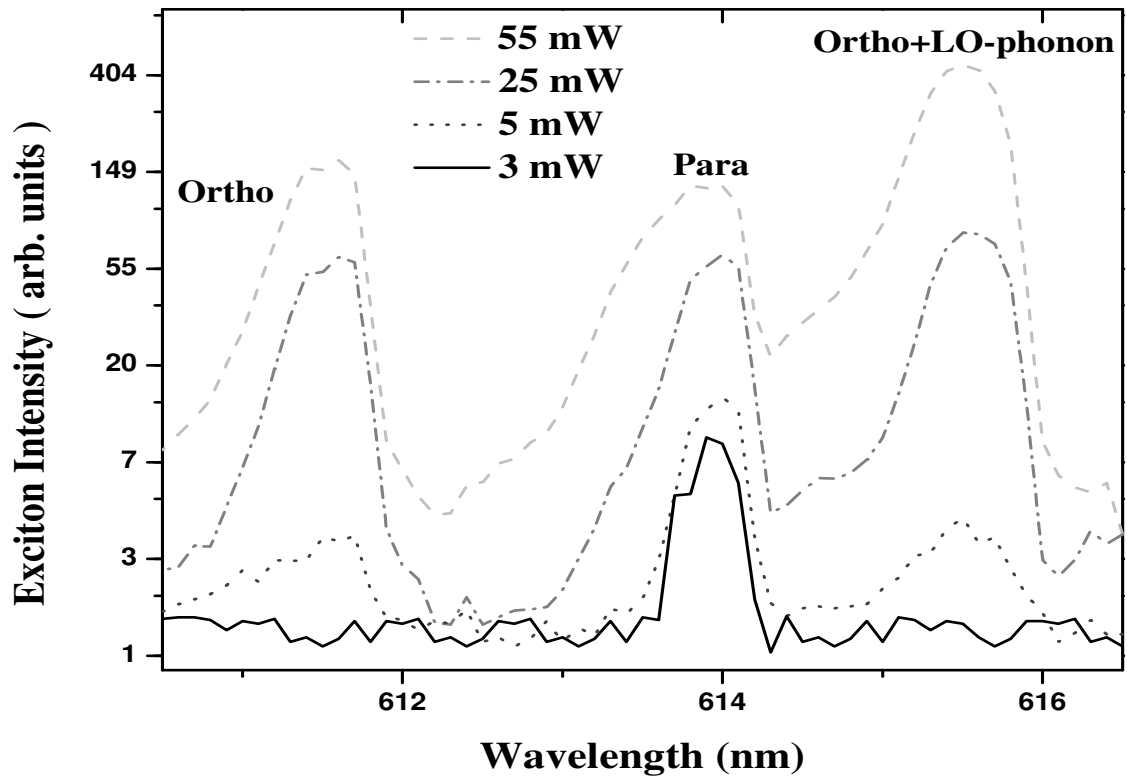}
\caption{\label{fig3.fig} Time-integrated exciton luminescence
intensity in the Cu$_2$O sample for several different laser
powers, at 1.6 Kelvin and under 1.9 kbar stress along the [001]
direction, with two-photon one-beam excitation resonant with the
paraexciton state.}
\end{center}
\end{figure}

\begin{figure}[!ht]
\begin{center}
\includegraphics[width=7in,height=5in]{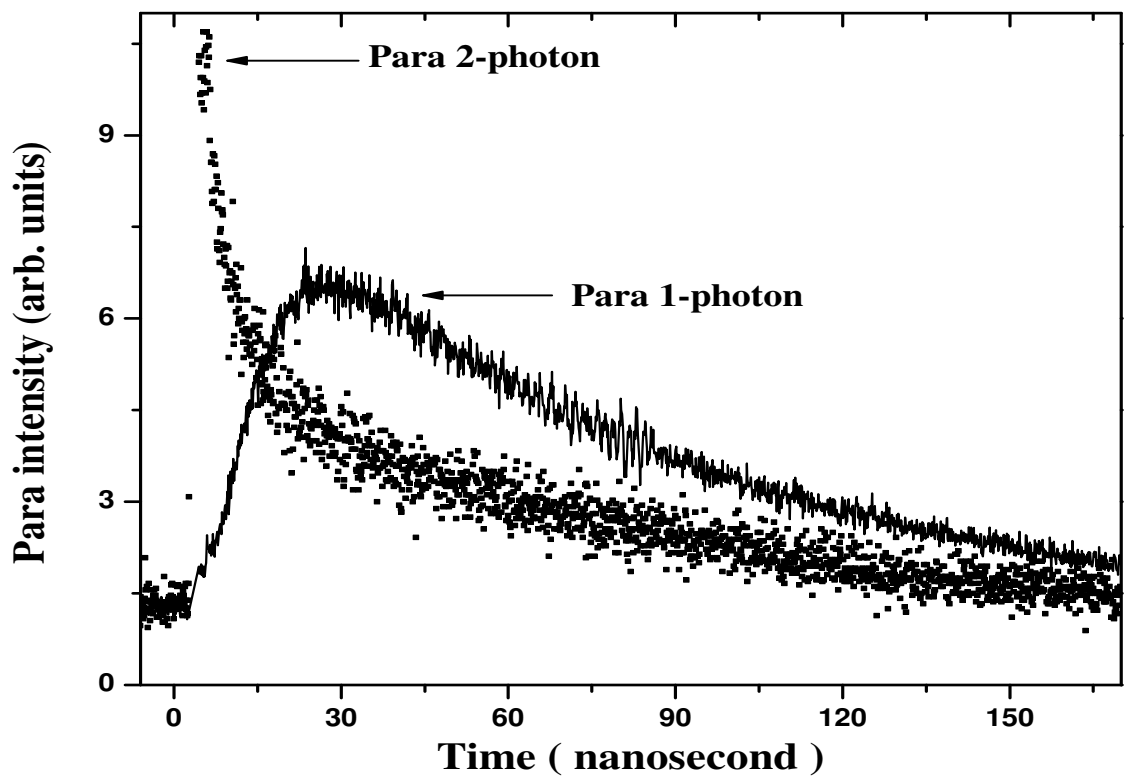}
\caption{\label{fig4.fig} Paraexciton luminescence as a function
of the time after the laser pulse, for a Cu$_2$O sample at 1.6
Kelvin and under 1.9 kbar stress along the [001] direction. Black
dots: two-photon excitation resonant with the paraexciton state.
Black solid line: single-photon excitation tuned to the bottom of
the orthoexciton phonon-assisted absorption. The power of the red
and the IR laser were adjusted to give the same total integrated
exciton luminescence intensity.}
\end{center}
\end{figure}

\begin{figure}[!ht]
\begin{center}
\includegraphics[width=7in,height=5in]{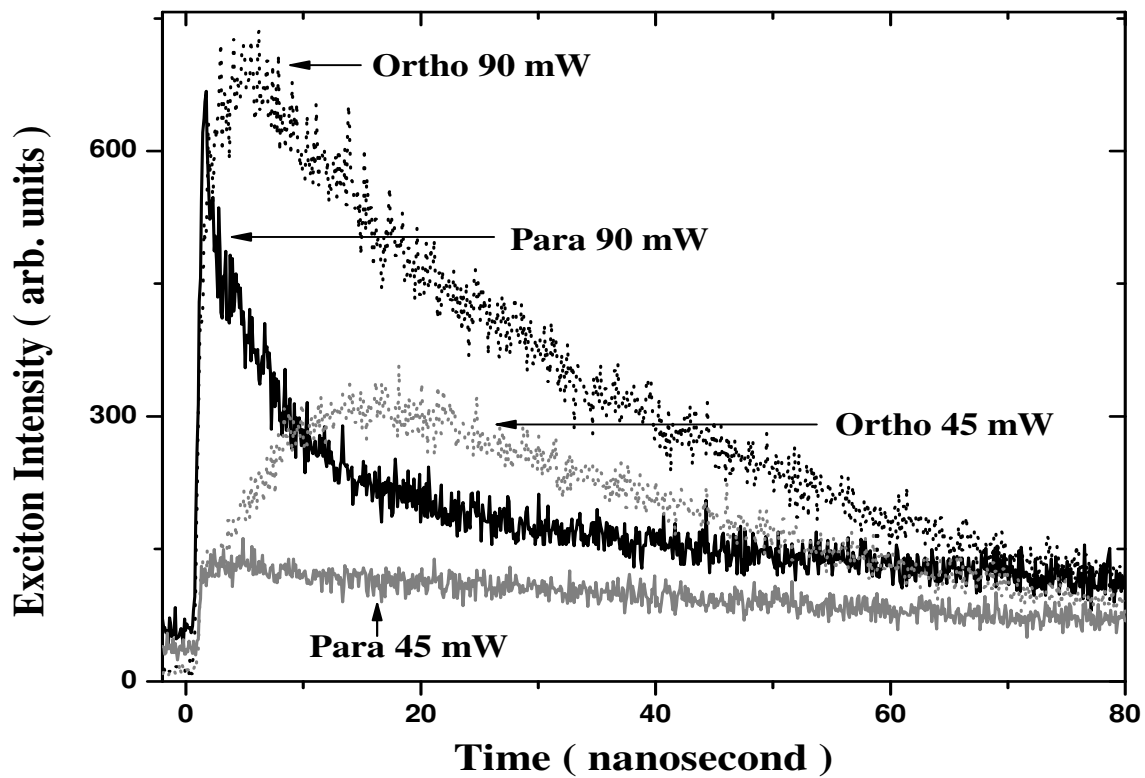}
\caption{\label{fig5.fig} Time-resolved paraexciton and
orthoexciton luminescence in the Cu$_2$O sample for two different
laser powers, at 1.6 Kelvin and under 1.9 kbar stress along the
[001] axis, with two-photon excitation resonant with the
paraexciton state. The black dots and black solid line are the
orthoexciton and paraexciton luminescence intensity at laser power
of 90 mW, respectively,  and the gray dots and gray solid line are
the orthoexciton and paraexciton luminescence intensity at laser
power of 45 mW, respectively.}
\end{center}
\end{figure}

\begin{figure}[!ht]
\begin{center}
\includegraphics[width=7in,height=5in]{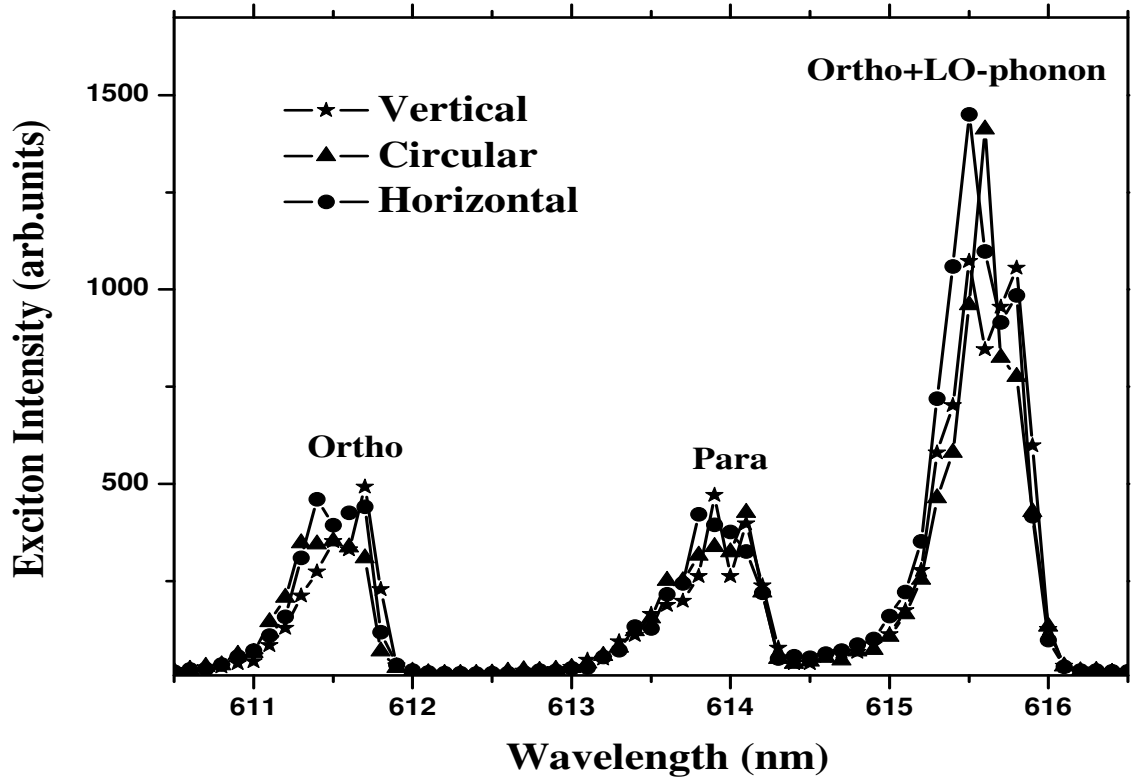}
\caption{\label{fig6.fig} Polarization dependence for the total
integrated paraexciton and orthoexciton luminescence intensities
for a Cu$_2$O sample at 1.6 Kelvin under 1.9 kbar stress along the
[001] direction, with two-photon excitation resonant with the
paraexciton state.}
\end{center}
\end{figure}

\begin{figure}[!ht]
\begin{center}
\includegraphics[width=7in,height=5in]{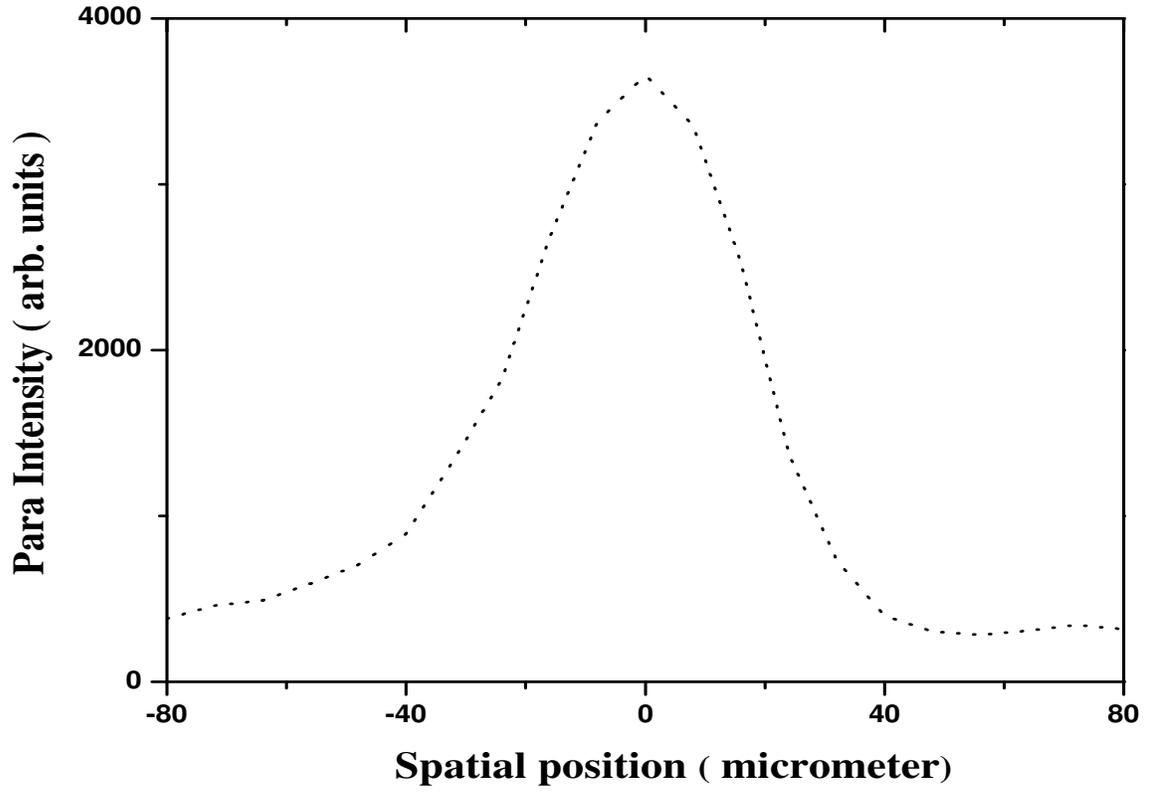}
\caption{\label{fig7.fig} Spatial profile of the paraexciton
luminescence in the stress-induced harmonic potenial trap for the
excitons, for the Cu$_2$O sample at 1.6 Kelvin under 1.9 kbar
stress along the [001] direction, with two-photon excitations
resonant with the paraexciton state.}
\end{center}
\end{figure}

\begin{figure}[!ht]
\begin{center}
\includegraphics[width=7in,height=5in]{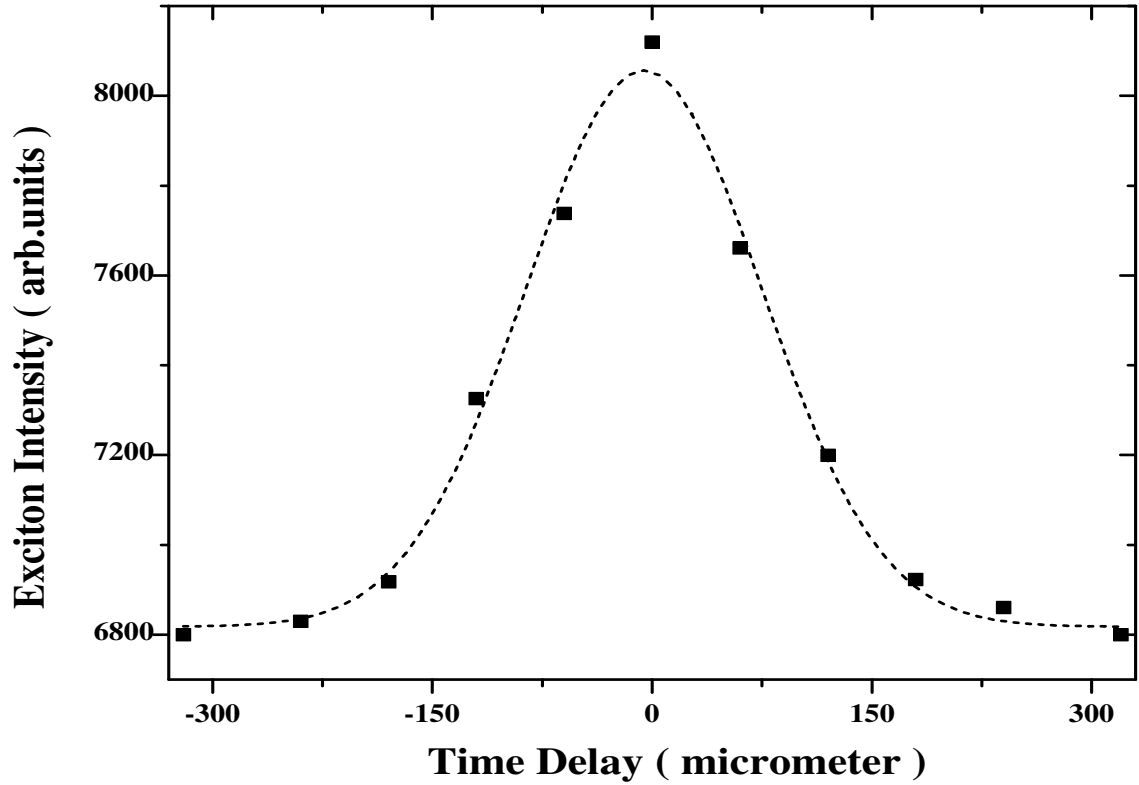}
\caption{\label{fig8.fig} Total integrated exciton luminescence
intensity as a function of the time delay between two laser pulses
for two-beam, two-photon excitation resonant with the paraexciton
state, for the Cu$_2$O sample at 1.6 Kelvin under 1.9 kbar stress
along the [001] direction. Black squares: experimental data.
Dashed line: Gaussian fit.}
\end{center}
\end{figure}

\begin{figure}[!ht]
\begin{center}
\includegraphics[width=7in,height=5in]{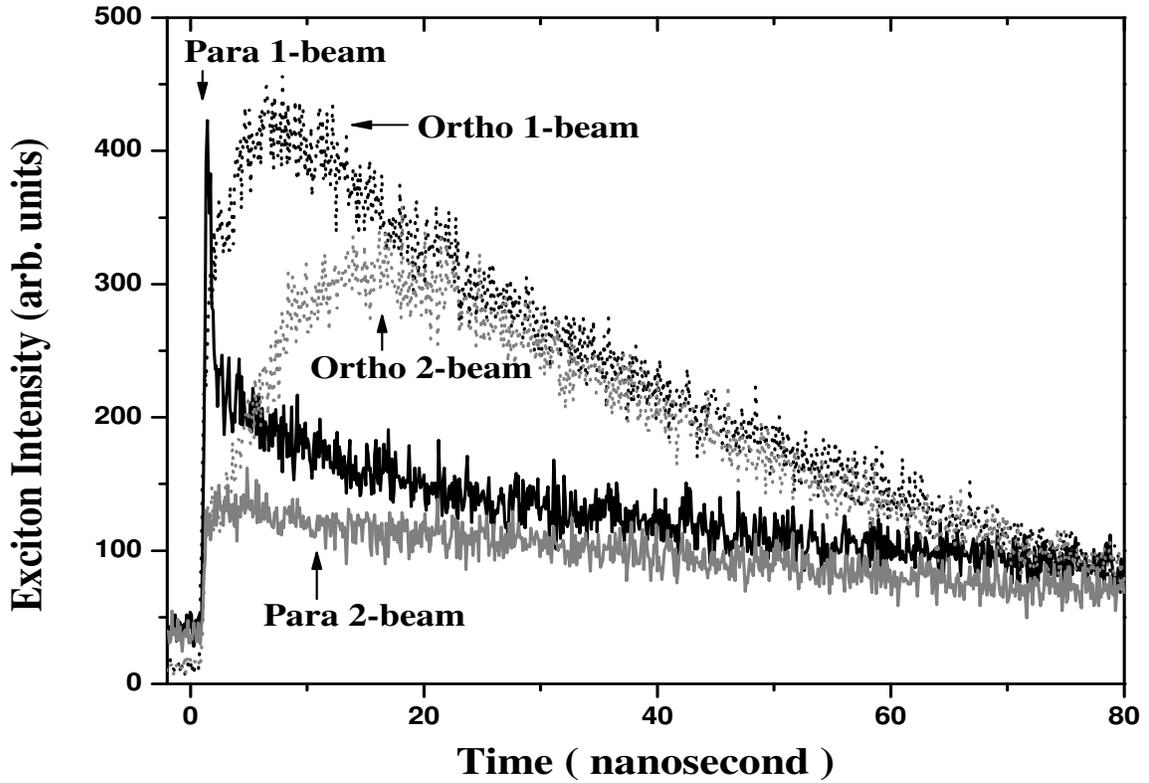}
\caption{\label{fig9.fig} Exciton luminescence intensity as a
function of the time after the laser pulses for one-beam and
two-beam two-photon excitation resonant with the paraexciton
state, with the same total laser power, for the Cu$_2$O sample at
1.6 Kelvin under 1.9 kbar stress along the [001] direction. The
black dots and and the black solid line are orthoexciton and
paraexciton luminescence intensity, respectively, for one-beam
excitation, and the gray dots and gray solid line are the
orthoexciton and paraexciton luminescence intensity, respectively,
for the two-beam excitation.}
\end{center}
\end{figure}

\end {document}